\documentclass{jps-cp}
\usepackage{color}
\usepackage{amsmath, amssymb,bm}
\usepackage{comment}

\usepackage{txfonts} 

\title{Reflection-Symmetry Protected Antiferromagnetic Topological Insulator in Three-Dimensional Heavy-Fermion Systems}
\author{Kazuhiro \textsc{Kimura}$^{1}$, Tsuneya \textsc{Yoshida}$^{2}$ and Norio \textsc{Kawakami}$^{1}$}
\inst{$^{1}$Department of Physics, Kyoto University, Kyoto 606-8502, Japan \\
$^{2}$Department of Physics, University of Tsukuba, Tsukuba, Ibaraki 305-8571, Japan}
\email{kimura.kazuhiro.85n@st.kyoto-u.ac.jp}

\recdate{\today}

\abst{We study the topological properties of an antiferromagnetic phase with reflection symmetry in three-dimensional heavy-fermion systems. 
We here propose a reflection-symmetric topological state in the three-dimensional antiferromagnetic phase and demonstrate how the paramagnetic phase changes into the antiferromagnetic topological phase of $f$-electron materials such as SmB$_6$.
}
\kword{antiferromagnetic topological insulator, heavy-fermion system, reflection symmetry}

\begin{document}
\maketitle
\section{Introduction}
Nowadays, the notion of band topology \cite{hasan_rmp_2010,qi_rmp_2011} has become much more ubiquitous and the topological materials are of great interest in condensed matter physics. The characteristics of topological insulators (TIs) is insulating bulk and their metallic edge (surface) states which are protected by local symmetry. Furthermore, the notion of topology has been generalized to metallic (gapless) phases\cite{wan_prb_2011,burkov_prl_2011,yoshida_prb_2013_ssti} such as topological semimetals or generalized to phases with crystalline (non-local) symmetry such as topological crystalline insulators\cite{fu_prl_2011,teo_prb_2008}. The topological properties of materials have been of great interest from the viewpoint of theoretical and experimental investigations.

Recently, the effects of Coulomb interaction on topological phases have attracted much attention because of nontrivial features that do not emerge in weakly correlated systems\cite{rachel_prog_2018,pesin_nat_2010,yoshida_prb_2012,yoshida_prl_2014}. 
For example, topological Kondo insulator\cite{dzero_prl_2010,dzero_review_2016,peters_prb_2016}, magnetic orders in topological insulating phases, and the change of topological classification\cite{fidkowski_prb_2010,fidkowski_prb_2011}, etc.
In the presence of the strong correlation effects, topological states become much more enriched and interesting because of an interplay between topology and correlation.

In previous studies, focusing on the topological properties of antiferromagnetic (AF) phases\cite{mong_prb_2010,yoshida_prb_2013}, an AF topological insulator (AFTI) has been proposed for three-dimensional (3D) systems, which is referred to as a $Z_2$-AFTIs 
characterized by a $Z_2$ topological number.
In such AFTIs, the AF order breaks time-reversal ($\Theta$) and primitive-lattice translation ($T_{1/2}$) symmetries but preserves their combined symmetry $S=T_{1/2} \Theta$. Under this symmetry, the system has a $Z_2$ topological number, which is related to a strong index of 3D topological insulators with time-reversal symmetry. Note that this $Z_2$ number is allowed only for 3D systems. 
Recently, it is found that the van der Waals layered compound MnBi$_2$Te$_4$\cite{kimura_afti_mnbite} could be a possible candidate for $Z_2$-AFTI.

In $f$-electron materials, SmB$_6$ and YbB$_{12}$ are good candidates of topological Kondo insulator, providing a good platform to study an interplay of Coulomb interaction and band topology.
Although SmB$_6$\cite{takimoto_jpsj_2011,lu_prl_2013} has been known as a Kondo insulator for about 40 years, it has been a long-standing problem about the saturation of its electrical resistivity below 4 K.
It has been recently proposed that the saturation originates from the topological surface state and this has been confirmed by angle-resolved photoemission spectroscopy measurements\cite{xu_ncom_2014,neupane_ncom_2013}.
More recently, another aspect of SmB$_6$ has been addressed experimentally.
A non-magnetic topological Kondo insulator, SmB$_6$, at ambient pressure turns into a metallic AF phase with increasing the pressure above 6GPa \cite{barla_orl_2005,derr_prb_2008,nishiyama_jpsj_2013,butch_prl_2016,zhou_sbull_2017}, but its magnetic configuration is not yet clarified experimentally. 
The topological properties of the AF phase have attracted much attention from the viewpoint of theoretical and experimental investigation.
From the first-principle calculation, it is pointed out that the pressurized SmB$_6$ would serve as a better candidate for a $Z_2$-AFTI and the A-type AF (A-AF) configuration is proposed for its ground state \cite{kw_chang_prb_2018}.

In this paper, we focus on the topological properties in AF phases from the viewpoint of crystalline symmetry in addition to the above $Z_2$-AFTI.
To be specific, we here consider it in the 3D heavy-fermion systems 
and demonstrate how the paramagnetic (PM) phase changes into the AF phase by using the Hartree-Fock (HF) approximation. 
We focus on the A-AF phase of SmB$_6$ which is proposed by the first-principle study\cite{kw_chang_prb_2018}.
In the previous study, we proposed a reflection-symmetric AFTI phase in 2D\cite{kimura_jpsj_2018}, which can have a topologically nontrivial structure specified by a mirror Chern number.
We here demonstrate a reflection-symmetric AFTI phase in 3D and point out the possibility of material realization in pressurized SmB$_6$.

The rest of paper is organized as follows. In Sec. \ref{model}, we describe our setup and give a brief explanation of our approach. 
In Sec. \ref{result}, we study the topological properties of the AF phase.
The last section \ref{summary} is devoted to a brief summary.

\section{Model and Methods\label{model}}

We consider the following periodic Anderson model with nonlocal $d$-$f$ hybridization in the orthorhombic crystal in Fig. \ref{fig:fig1}(a),
\begin{subequations}
\label{eq:Hamiltonian}
\begin{eqnarray}
H&=&\sum_{\bm{k}}
\begin{pmatrix}
\bm{d}^{\dagger}_{\bm{k}}  &\bm{f}^{\dagger}_{\bm{k}}
\end{pmatrix}
\begin{pmatrix}
\epsilon^d_{\bm{k}}   &V_{\bm{k}}  \\
V^{\dagger}_{\bm{k}}&\epsilon^f_{\bm{k}} 
\end{pmatrix}
\begin{pmatrix}
\bm{d}_{\bm{k}}  \\
\bm{f}_{\bm{k}}
\end{pmatrix}
+\sum_{j}U_{f}n^{f}_{j\uparrow}n^{f}_{j{\downarrow}}, 
\end{eqnarray}
with
\begin{eqnarray}
\epsilon^d_{\bm{k}} &=& 
[-2t_d(\cos{k_x}+\cos{k_y})-4t_d^{\prime}(\cos{k_x}\cos{k_y})]\tau_0 \sigma_0  \nonumber \\
&&+[-2t_d\cos{k_z}-4t_d^{\prime}\cos{k_z}(\cos{k_x}+\cos{k_y})]\tau_x \sigma_0,   
\\
\epsilon_{\bm{k}}^f &=& 
[\epsilon_f-2t_f(\cos{k_x}+\cos{k_y})-4t_f^{\prime}(\cos{k_x}\cos{k_y})]\tau_0 \sigma_0  \nonumber \\
&&+[-2t_f\cos{k_z}-4t_f^{\prime}\cos{k_z}(\cos{k_x}+\cos{k_y})]\tau_x \sigma_0,   \\
V_{\bm{k}}&=& 
-2V [\sin{k_z}\tau_x \sigma_z +\sin{k_x} \tau_0\sigma_x +\sin{k_y} \tau_0 \sigma_y], 
\end{eqnarray}
\end{subequations}
where $\epsilon^d_{\bm{k}}$ ($\epsilon^f_{\bm{k}}$) is the dispersion of $d$- ($f$-) electrons, 
$V_{\bm{k}}$ is a Fourier component of the nonlocal $d$-$f$ hybridization, 
$n_{i\sigma}^{f}$ is the number operator, 
$\bm{k}$ is a wave number, and $\sigma_{i}$ and $\tau_{i}$ ($i =0,x,y,z$) are the Pauli matrices for spins and sublattices.
The annihilation operators are defined as
$\bm{d}_{\bm{k}}=
\begin{pmatrix}
d_{\bm{k}\uparrow}^A  &
d_{\bm{k}\uparrow}^B  &
d_{\bm{k}\downarrow}^A  &
d_{\bm{k}\downarrow}^B
\end{pmatrix}
^{T}$ and $\bm{f}_{\bm{k}}=
\begin{pmatrix}
f_{\bm{k}\uparrow}^A  &
f_{\bm{k}\uparrow}^B  &
f_{\bm{k}\downarrow}^A  &
f_{\bm{k}\downarrow}^B
\end{pmatrix}
^{T}$.
Here, $t_d$, $t_f$, $t_d^{\prime}$, and $t_f^{\prime}$ are hopping parameters, 
$V$ denotes the $d$-$f$ hybridization, and
$\epsilon_f$ is the difference between the $d$- and $f$-electron energies.
The topological properties stem from nonlocal $d$-$f$ hybridization\cite{dzero_prl_2010,dzero_review_2016}
 which reflects spin-orbit coupling.

In order to study the topological properties of the AF phase in Eq. (\ref{eq:Hamiltonian}), we employ the following HF approximation for the Hubbard interaction term:
$n_{i \uparrow}^f n_{i \downarrow}^f\sim \langle n_{i \downarrow}^f \rangle n_{i \uparrow} ^f+ \langle n_{i \uparrow}^f \rangle n_{i \downarrow}^f-  \langle  s_{i-}^f \rangle  s_{i-}^f -\langle s_{i+}^f \rangle  s_{i+}^f$,
where $s_{i-}^f:=f_{i\downarrow}^{\dagger}f_{i\uparrow}$, $s_{i+}^f:=f_{i\uparrow}^{\dagger}f_{i\downarrow}$.
Here, $\langle \cdots \rangle$ denotes the expectation value at zero temperature.
Thus, the mean-field Hamiltonian is given by 
\begin{eqnarray}
\mathcal{H}^{\rm mf}(\bm{k})&=&
\begin{pmatrix}
\epsilon^d_{\bm{k}}   &V_{\bm{k}}  \\
V^{\dagger}_{\bm{k}}&\epsilon^f_{\bm{k}}+h^{f}_{int}
\end{pmatrix}
, {\rm with} \, \,   
h^{f}_{int}=U_{f}
\begin{pmatrix}
 \langle n^{f A}_{0\downarrow} \rangle  &0 &-\langle s^{f A}_{0+} \rangle  &0 \\
0 & \langle n^{f B}_{0\downarrow} \rangle &0 &- \langle s^{f B}_{0+} \rangle \\
-\langle s^{f A}_{0-} \rangle &0 &\langle n^{f A}_{0\uparrow}\rangle  &0 \\
0 &-\langle s^{f B}_{0-} \rangle&0 &\langle n^{f B}_{0\uparrow} \rangle
\end{pmatrix}
.
\end{eqnarray}

\section{Topological Properties of Antiferromagnetic Phases \label{result}}
The non-magnetic topological Kondo insulator SmB$_6$ at ambient pressure turns into a metallic AF phase by high pressure above 6GPa, 
although its magnetic ordering pattern is not clarified experimentally.
In a previous theoretical study\cite{kw_chang_prb_2018}, it is pointed out that the ground state is the A-AF phase with $\vec{m}\parallel \hat{z}$ and 
the $Z_2$-AFTI phase emerges in this AF phase.
In this paper, motivated by recent experimental studies in pressurized SmB$_6$, 
we consider the topological properties of the AF phase from the viewpoint of reflection symmetry. 
In our study, we consider the A-AF phase as shown in Fig. \ref{fig:fig1}(a) with $\vec{m}\parallel \hat{z}$ and $\vec{m}\parallel \hat{x}$.
The model we employ here is a topological mirror Kondo insulator which was previously used to address a nonmagnetic cubic Kondo insulating phase \cite{legner_prb_2014} with band inversion at X-point\cite{takimoto_jpsj_2011,lu_prl_2013} which is a two-orbital effective model of SmB$_6$ in its PM phase.
We elucidate below that the system can change form a PM phase to an AF phase without breaking its reflection symmetry, thus leading to an AFTI protected by the reflection symmetry.

\subsection{Definition of topological invariants}
In order to consider the magnetically ordered topological insulating states with a reflection symmetry, we introduce the corresponding topological number.
In the reflection-symmetric plane in 3D Brillouin zone (BZ), all the eigenstates are characterized by their reflection parities and divided into two subspaces as
\begin{eqnarray}
\mathcal{H}^{\rm mf}(\bm{k})=
\begin{pmatrix}
{\mathcal H}_{ M=+i}(\bm{k}) &0 \\
0&{\mathcal H}_{M=-i}(\bm{k}) 
\end{pmatrix}
,
\end{eqnarray}
where $M$ means the reflection operator with $M^2=-1$ and its eigenvalues are $M=\pm i$.
The Chern numbers of each reflection subspace $C_{M=\pm i}$ can be defined on the 2D subspace of 3D BZ, like $k_z=0$ plane and it is given as
\begin{eqnarray}
C_{M=\pm i}  &=& \frac{1}{2\pi} \sum_{i }\int_{S} [\nabla_{\bm{k}}\times \mathcal{A}_{i, M}(\bm{k})]_z dk_x dk_y,
\end{eqnarray}
where $\mathcal{A}_{i, M}(\bm{k})=-i\langle u_{i,M}(\bm{k})|\nabla_{\bm{k}}|u_{i,M}(\bm{k})\rangle$ is the Berry connection,
 where $|u_{i,M}(\bm{k}) \rangle$ is a Bloch state with occupied band index $i$, which is an eigenstate of ${\cal H}_{M}(\bm{k})$.
The net Chern number $C_{net}=C_{M=+i}+C_{M=-i}$ and the mirror Chern number $C_m=(C_{M=+i}-C_{M=-i})/2$ are defined by  $C_{M=\pm i}$.

\begin{figure}[h]
\begin{center}
\includegraphics[width=0.8\hsize]{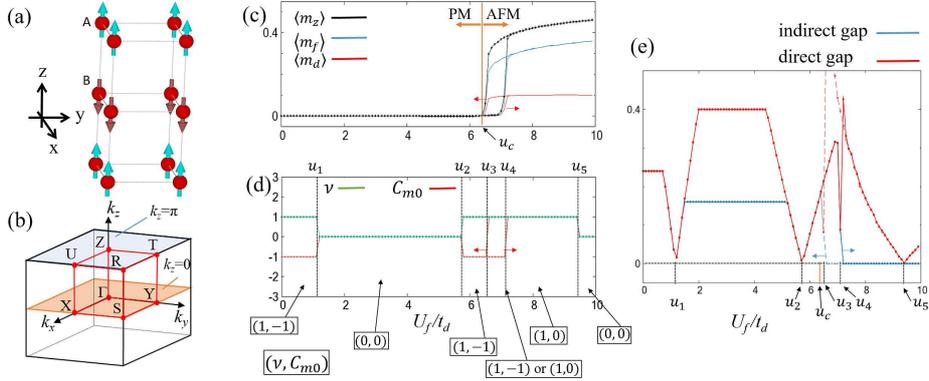}
\end{center}
\caption{
(Color online) Magnetic and topological properties for $V=0.8$, obtained with the HF approximation: 
(a) spin configuration of the A-AF phase with $\vec{m}\parallel \hat{z}$, 
(b) BZ of orthorhombic crystal, 
(c) staggered magnetic moment,
(d) mirror Chern number $C_{m0}$ on $k_z=0$ plane and $Z_2$ number $\nu$,  
(e) indirect gap and direct gap.
In (a), we choose the $z$-direction as the doubled axis.
In (c), a hysteresis loop appears because of the first-order phase transition.
In (e), the indirect gap is the band gap between the conduction and valence bands ${\rm min}(E^{\rm conduction}_{\bm{k}}-E^{\rm valence}_{\bm{k}^{\prime}})$ ($\bm{k}$ is not necessarily equal to $\bm{k}^{\prime}$) while the direct gap is the band gap at wave number $\bm{k}$ ${\rm min}(E^{\rm conduction}_{\bm{k}}-E^{\rm valence}_{\bm{k}})$, where indirect gap $\leq$ direct gap.
The region where the indirect gap is closed is semimetallic, and the points where the direct gap is closed denote the topological phase transition.
}
\label{fig:fig1}
\end{figure}

Next, we discuss the $Z_2$ invariant of the AF phases \cite{mong_prb_2010} with inversion symmetry.
In the collinear AF phase, the AF order breaks time-reversal ($\Theta$) and primitive-lattice translational ($T_{1/2}$) symmetries but preserves their combined symmetry $S=T_{1/2}\Theta$. 
The topological invariant of this AF phase with inversion symmetry is given by a $Z_2$ number $\nu (=0, 1)$ as follows, $(-1)^{\nu} = \prod_{\bm{k}_m}\delta_m$,
where $\bm{k}_m$ are four Kramers degenerate momenta and $\delta_m=\prod_i \xi_i(\bm{k_m})$ are products of the parity $\xi_i(\bm{k_m})$ of the occupied band $i$. 
At the Kramers degenerate momenta, the parity of $d$- ($f$-) orbital is given by $\xi_i(\bm{k_m})=+1 (-1)$. 
We note, in A-AF phase with the doubled $z$-axis, that the $Z_2$ number can be defined by the product of each parity at ${\rm X}$, ${\rm S}$, ${\rm Y}$, and $\Gamma$ in Fig. \ref{fig:fig1}(b). 

\subsection{Numerical results}
First, we here discuss the obtained results for the A-AF phase with $\vec{m}\parallel \hat{z}$ at half-filling as shown in Fig. \ref{fig:fig1}(a).
In this case, the independent reflection plane is only $k_z=0$ where the reflection operator $M$ inverts the $z$-axis, thus we consider the mirror Chern number $C_{m0}$ on $k_z=0$ plane.
The values of the parameters we employ in the following are $t_d=1$ (energy unit), $t_d^{\prime}=-0.4$, $t_f=-0.1$, $t_f^{\prime}=0.04$, $\epsilon_f=-2$, $V=0.8$.
In this parameter set, the non-interacting Hamiltonian realizes the strong topological insulator with band inversion at X-point\cite{legner_prb_2014,legner_prl_2015}.
The magnetic properties are studied by the HF method and the Chern number is calculated by 
the method proposed in Ref. \cite{hukui_jpsj_2005}, which is efficient for numerical calculations.

We obtain some AF topological (AFT) phases for $V=0.8$ as shown in Figs.\ref{fig:fig1}(c)-(e).
Note that the results are not sensitive to the value of $V$ and we discuss it later.
With increasing the interaction $U_f$, the PM phase changes into the AF phase at $U_f=u_c$ with a large hysteresis loop in Figs.\ref{fig:fig1}(c) and a magnetic configuration in Fig. \ref{fig:fig1}(a).  
Figure \ref{fig:fig1}(d) indicates that in the weak interacting and PM region $U_f<u_1$, the topological insulating state is characterized by non-trivial topological numbers $(\nu, C_{m0})$.
In the strongly interacting and AF region $u_c<U_f<u_5$, the AFT phase which is characterized by the $Z_2$ number and the mirror Chern $C_{m0}$ numbers emerges.
In the hysteresis region, the mirror Chern has a hysteresis loop which is induced by the change of the band structure in the AF phase. 
The above topological phase transition points are consistent with the gap closing point of direct gap in Fig. \ref{fig:fig1}(e). 
In the AF phase, the system becomes metallic where the indirect gap is closed in $U_f>u_4$ (or $u_3$) in Fig. \ref{fig:fig1}(e).   
Note, however, that the topological properties still remain intact in this region because the direct gap is not closed.
The topological numbers are still well-defined in the region where the direct gap is open.
We confirm an aspect of the reflection symmetry in an AFT semetallic phase characterized by a mirror Chern number and a $Z_2$ number.

\begin{figure}[h]
\begin{center}
\includegraphics[width=0.8\hsize]{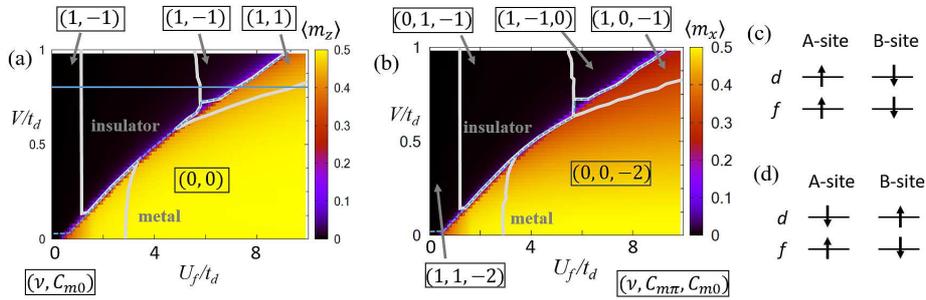}
\end{center}
\caption{
(Color online) (a)[(b)]: Phase diagram of AFT semimetal with $\vec{m}\parallel \hat{z}$ [ $\vec{m}\parallel \hat{x}$ ] as functions of the interaction $U_f$ and hybridization $V$. 
(c)[(d)]: Spin configuration of the AF phase for panel (a)[(b)].
The gray (blue dashed) line denotes the topological (insulator-metal) phase transition line.
In the metallic region, the indirect gap is closed; thus, there is a Fermi surface. However, the direct gap is not closed; thus the some topological numbers are still well-defined. 
$\nu$ means the $Z_2$ number on $k_z=0$ plane.
In (a),  MC$_0$ means the mirror Chern number on $k_z=0$ plane.
In (b), MC$_0$ and MC$_\pi$ mean the mirror Chern number on $k_x=0$ and $k_x=\pi$ plane.
The blue line represents the $V=0.8$
}
\label{fig:fig2}
\end{figure}

Summarizing all these results with $\vec{m}\parallel \hat{z}$, we obtain the phase diagram shown in Fig. \ref{fig:fig2}(a). 
We also study the case of $\vec{m}\parallel \hat{x}$ in Fig. \ref{fig:fig2}(b).
The horizontal axis denotes the strength of the interaction $U_f$ and the vertical axis the strength of hybridization $V$.
The above analysis is done for $V=0.8$ on the blue line in Fig. \ref{fig:fig2}(a).
In the AF phase for these parameters, a semimetallic AFT phase is realized, which we refer to as an AFT semimetal.
The spin configuration in the AF phase is shown in Fig. \ref{fig:fig2}(c), where the $d$- and $f$-magnetization are pointed in the same direction on the same site.
The mirror Chern number and the $Z_2$ number have various values in the phase diagram, which are changed by the shift of the $f$-band.

In addition, we discuss another case with $\vec{m}\parallel \hat{x}$ in Fig. \ref{fig:fig2}(b).
In this case, the independent reflection planes are $k_x=0$ and $k_x=\pi$ where the reflection operator $M$ flips the $x$-axis, thus we consider the mirror Chern numbers $C_{m0}$ and $C_{m\pi}$ on $k_x=0$ and $k_x=\pi$ planes.
We obtain some AFT semimetallic phases.
A prominent feature in this case is that a reflection-symmetric AFT phase emerges, which is characterized by only non-zero mirror Chern number $(0, 0, -2)$ and has never been discussed in the previous study.
The spin configuration in the AF phase is shown in Fig. \ref{fig:fig2}(d), where the $d$- and $f$-magnetization are aligned in the opposite direction on the same site.

\section{Summary and Discussion\label{summary}}
In this paper, we have analyzed the AFT phase in 3D by taking into account reflection symmetry.
Specifically, motivated by recent experimental studies of the magnetic metallic phase in pressurized SmB$_6$, 
we have elucidated the novel aspect of the reflection-symmetry protected AFT in two types A-AF phases.
Our numerical results have revealed the emergence of the 3D topological crystalline insulating states in AF phase for interacting systems such as A-AF phase in SmB$_6$.
In particular, we have shown the AFT states characterized by mirror Chern numbers in addition to $Z_2$ number.
In the case of $\vec{m}\parallel \hat{x}$, a reflection-symmetric AFT semimetallic phase appears.

We finish this paper with a comment on a related work. In Ref. \cite{kw_chang_prb_2018} a first principle calculation for SmB$_6$ has elucidated the surface excitation spectrum for the magnetic phase. In that article, it has been reported that at a reflection plane, there exist the gapless surface states which cannot be understand only with the magnetic translation symmetry (i.e., the product of the translation and the time-reversal operation).

We consider that our results elucidating the additional topological properties with reflection symmetry may explain the origin of these gapless states whose origin remains unclear. In order to understand it, detailed numerical calculations are required. We leave this issue as a future work.

This work was partly supported by JSPS KAKENHI Grant No. JP15H05855, JP18H01140, JP18H05842 and JP19H01838.
The numerical calculations were performed on the supercomputer at the Institute for Solid State Physics in the University of Tokyo, and SR16000 at Yukawa Institute for Theoretical Physics in Kyoto University.


\end{document}